# Gate Tunable Quantum Oscillations in Air-Stable and High Mobility Few-Layer Phosphorene Heterostructures

Nathaniel Gillgren[1], Darshana Wickramaratne[2], Yanmeng Shi[1], Tim Espiritu[1], Jiawei Yang[1], Jin Hu[3], Jiang Wei[3], Xue Liu[3], Zhiqiang Mao[3], Kenji Watanabe[4], Takashi Taniguchi[4], Marc Bockrath[1], Yafis Barlas[1,2], Roger K. Lake[2], Chun Ning Lau[1*]

[1] Department of Physics and Astronomy, University of California, Riverside, Riverside, CA 92521
[2] Department of Electrical and Computer Engineering, University of California, Riverside, Riverside, CA 92521
[3] Department of Physics and Engineering Physics, Tulane University, New Orleans, LA 70118
[4] National Institute for Materials Science, 1-1 Namiki Tsukuba Ibaraki 305-0044 Japan.

ABSTRACT
As the only non-carbon elemental layered allotrope, few-layer black phosphorus or phosphorene has emerged as a novel two-dimensional (2D) semiconductor with both high bulk mobility and a band gap. Here we report fabrication and transport measurements of phosphorene-hexagonal BN (hBN) heterostructures with one-dimensional (1D) edge contacts. These transistors are stable in ambient conditions for >300 hours, and display ambipolar behavior, a gate-dependent metal-insulator transition, and mobility up to 4000 $cm^2/Vs$. At low temperatures, we observe gate-tunable Shubnikov de Haas (SdH) magneto-oscillations and Zeeman splitting in magnetic field with an estimated $g$-factor ~2. The cyclotron mass of few-layer phosphorene holes is determined to increase from 0.25 to 0.31 $m_e$ as the Fermi level moves towards the valence band edge. Our results underscore the potential of few-layer phosphorene (FLP) as both a platform for novel 2D physics and an electronic material for semiconductor applications.

Phosphorene is single- or few-layers of black phosphorus[1-6] that is the most stable form of phosphorus. Apart from carbon, it is the only known element with a stable layered allotrope. In single-layer phosphorene, the atoms are arranged in a honeycomb structure, much like graphene, though the atoms are puckered (Fig. 1a). Like graphene, charge carriers in bulk black phosphorus can have exceedingly high mobility, >50,000 $cm^2/Vs$[7]; but unlike the gapless graphene band structure, bulk black phosphorus hosts a direct band gap. The size of the band gap is thickness dependent, and is predicted to vary from 0.35 eV in bulk to ~2 eV in monolayers[1, 3, 7-12]. Furthermore, it is also predicted to have unusual properties such as anisotropic transport[1, 3], large thermoelectric power[13-15], and a band gap that is tunable by strain[16, 17]. Recently field effect transistors based on few-layer phosphorene (FLP) have been demonstrated, with mobility ~300–1000 $cm^2/Vs$[2, 5, 18]. Thus, phosphorene is emerging as a new two-dimensional (2D) semiconductor with tremendous promise for electronics, thermal and optoelectronics applications, as well as a model system with interesting physical properties.

Despite the recent surge in interest in this new 2D material, several major challenges remain. For instance, when phosphorene is exposed to air or moisture it reacts to form phosphoric acid, which degrades or destroys the material[1, 19-21]. Therefore, in order to develop stable electronic and optoelectronic devices from this material it must be protected from ambient conditions. Another challenge is that though device mobility is high compared to other 2D materials, the highest reported value (~1000 cm$^2$/Vs for 10nm-thick phosphorene sheets[18]) is still much lower than that of bulk, which is ~60,000 cm$^2$/Vs for holes and ~20,000 for electrons. Thus device mobility has much room for improvement; high-mobility devices will also enable exploration of phenomena that are not otherwise possible, such as ballistic transistors, directional transport and spintronics applications.

---

[*] Email: lau@physics.ucr.edu

Here we address both challenges by fabricating hBN/few-layer phosphorene/hBN heterostructures, in which the phosphorene layers are contacted via 1D edge contacts[22]. Such encapsulated devices are air-stable, exhibiting minimal degradation after more than 300 hours under ambient conditions. Electrical measurement on these hBN/phosphorene/hBN heterostructures reveal ambipolar transport with an on/off ratio exceeding $10^5$ and mobility ~400 cm$^2$/Vs at room temperature. At low temperatures, device mobility increases to ~4000 cm$^2$/Vs. In magnetic field $B$>3.5T, gate-tunable SdH oscillations are observed, enabling us to extract the cyclotron mass of few-layer phosphorene at different Fermi energies, ~0.25 to 0.31 $m_e$, where $m_e$ is the rest mass of electrons. These values are in good agreement with those obtained from *ab initio* calculations. Finally, at $B$>8T, we observe a doubling of the SdH period, suggesting the emergence of Zeeman splitting. From the oscillations' temperature dependence, we estimate that the *g*-factor is ~2. Our results point the way to fabrication of stable, high mobility devices for phosphorene and other air-sensitive 2D materials, and underscore its potential as a new platform for quantum transport and applications in 2D semiconductors.

BP bulk crystals are synthesized using chemical vapor transport technique ([23], also see Methods), or purchased from Smart Elements. To fabricate the devices, we first exfoliate the bottom hBN layers onto Si/SiO$_2$ substrates; few-layer phosphorene sheets and top hBN layers are exfoliated onto separate PDMS stamps, which are then successively transferred using a standard dry-transfer technique[19] to create hBN/phosphorene/hBN heterostructures. These layer transfer procedures are carried out in an inert atmosphere in a glove box to minimize exposure to oxygen and moisture. The completed stacks are etched into Hall bar geometry with exposed phosphorene edges, and metal electrodes consisting of 10 nm of Cr and 100 nm of gold are deposited to achieve 1D edge contacts[22]. The Si/SiO$_2$ substrate serves as the back gate, and, if desired, a top gate electrode can be added to the stack. A schematic of the fabrication process is shown in Fig. 1b, and a false-color optical image of the completed device in Fig. 1c. We note that this is the first report of successful 1D contacts to a 2D semiconductor.

Few-layer phosphorene is known to be unstable in air, and reacts to form phosphoric acid in a matter of hours[1, 19-21]. For standard phosphorene devices on SiO$_2$ substrates without encapsulation, both the device conductance and mobility degrade significantly within 24 hours. To test the stability of hBN-encapsulated phosphorene devices, we monitor the two-terminal conductance $G$ *vs.* back gate voltage $V_{bg}$ for such a device that is ~10 nm thick. The red curve in Fig. 1d displays $G(V_g)$ measured immediately after fabrication; ambipolar transport is observed, with the charge neutrality point at $V_g$=-3V, and a field-effect mobility of ~ 30 cm$^2$/Vs. The device is kept in ambient conditions in a drawer and monitored after 24, 48, 72, 120, 192 and 312 hours. At the end of the period, the charge neutrality point shifts to 1V, suggesting a small increase in electron doping; the device conductance and mobility decreases only slightly. Such stability over nearly a fortnight constitutes enormous improvement over "bare" phosphorene samples, and is in fact better than most conventional graphene devices that are chemically stable and inert. Thus, with further optimization, phosphorene may be realistically employed for electronic and optoelectronic applications.

Apart from providing a capping layer that protects phosphorene from oxygen and moisture, hBN also serves as a substrate that, because of its atomically flat surfaces and absence of dangling bonds, enables high mobility transport[24, 25]. Here we present data from a ~10 nm-thick hBN-encapsulated phosphorene device. Fig. 2a-b presents the field-effect transistor behavior $G(V_g)$ at room temperature and low temperature, respectively. At temperature $T$=300K, the device exhibits ambipolar transport, an on/off ratio >$10^5$, sub-threshold swing of ~100 mV/decade in the hole regime, and hole mobility of ~ 400 cm$^2$/Vs (Fig. 2a). Unlike the "bare" phosphorene devices[2], the $G(V_g)$ curves display minimal hysteresis, again underscoring device stability. At low temperature, the hole mobility increases to ~4000 cm$^2$/Vs at $T$=1.5K (Fig. 2b). We note that, *apart from graphene, this is the highest mobility value reported for 2D materials to date.* Its current-voltage characteristics in the hole-doped regime remain linear at all temperatures (Fig. 2c), indicating ohmic contacts.

To further explore transport in the few-layer phosphorene device, we explore its conductance at different gate voltages as temperature varies. For highly hole-doped regime ($V_g$<-30V), the four-terminal longitudinal resistance $R_{xx}$ decreases with decreasing temperature, indicating metallic behavior. However, as the Fermi level is tuned towards the band edge, *i.e.* for $V_g$>-25V, $R_{xx}$ increases drastically as $T$ is lowered, characteristic of an insulator (Fig. 2d). Fig. 2e plots $R_{xx}(T)$ for $V_g$=-70, -50, -30, -25, -20, -17 and -15V, respectively, where the clear dichotomy of gate-dependent metal-insulator transition is evident.

Further information on scattering mechanisms in the few-layer phosphorene device can be gleaned from the temperature dependence of mobility $\mu=\sigma/ne$. Here $\sigma$ is the conductivity of the device, $e$ the electron charge and $n$ the charge density. $n$ can be extracted from geometrical considerations as well as magneto-transport data (see discussion below). Fig. 2f displays $\mu(T)$ for 3 different $V_g$ values. When the Fermi level is deep in the valence band, $\mu$ increases with decreasing $T$ for $T$>70K, but saturates at lower temperatures. The initial enhancement of $\mu$ is expected from phonon-limited scattering, where $\mu \sim T^{-\alpha}$. For atomically thin 2D materials, the exponent $\alpha$ is predicted to be ~1.69 for MoS$_2$[26], and between 1 to 6 for graphene [27-32]. The saturation of $\mu$ at lower temperatures suggests impurity-dominated scattering. When the Fermi level moves closer to the valence band edge ($V_g$>-25V), $\mu$ decreases monotonically with $T$; this behavior is likely due to reduced screening and enhanced scattering from charged impurities at diminished doping level. Further experimental and theoretical efforts will be necessary to ascertain the scattering mechanisms at different temperature and doping regimes.

We now focus on transport behavior of the few-layer phosphorene device in a perpendicular magnetic field. Fig. 3a plots $\Delta R_{xx}$, in which a smooth background is subtracted from the longitudinal signal, as a function of $V_g$ (vertical axis) and $B$ (horizontal axis). Striking patterns of Shubnikov-de Haas (SdH) oscillations, appearing along straight lines that radiate from the charge neutrality point and $B$=0, are observable for $B$>3T. The charge neutrality point (or the center of the band gap) is extrapolated to be $V_g^{CNP}$~28V at $T$=1.5K. These quantum oscillations arise from the Landau quantization of cyclotron motion of charge carriers, and are often employed as a powerful tool to map Fermi surfaces of metals and semiconductors. Quantitatively, the oscillations are described by the Lifshitz-Kosevich formula for 2D systems[33, 34]

$$\frac{\Delta R_{xx}}{R_{xx}} \propto \frac{\lambda}{\sinh \lambda} e^{-\lambda_D} \cos\left(\frac{2\pi E_F}{\hbar \omega_c} + \pi\right) \qquad (1)$$

Here $\lambda = \frac{2\pi^2 k_B T}{\hbar \omega_c}$, $\omega_c = eB/m^*$ is the cyclotron frequency, $m^*$ the cyclotron mass of charge carriers, $k_B$ the Boltzmann's constant, $E_F$ the Fermi level and $\lambda_D = \frac{2\pi^2 k_B T_D}{\hbar \omega_c}$. $T_D$ is the Dingle temperature, given by $k_B T_D = \frac{\hbar}{2\pi\tau}$, where $\tau$ is the relaxation time of charge carriers. In 2D systems with spin degeneracy, $\frac{2\pi E_F}{\hbar \omega_c} = 2\pi \frac{nh}{2Be}$, regardless of the details of the dispersion relation; thus the oscillations in resistance are periodic in $nh/2Be$, independent of $m^*$. The amplitudes of the oscillations are exponentially dependent on $m^*$ and temperature. Fig. 3b displays line traces $\Delta R_{xx}(V_g)$ at constant $B$=2, 5, 8, 10 and 12T, where the oscillations are periodic in $V_g$. Fig. 3c plots $\Delta R_{xx}$ vs. $B$ (left panel) and $1/B$ (right panel) at constant $V_g$=-30, -40 and -60V, respectively. As expected from Eq. (1), the oscillations grow in amplitude as the Fermi level moves towards the band edge, and the period is given by $1/B_F=2e/nh$. Using the oscillation data, we determine the back gate coupling efficiency to be ~ $8.0 \times 10^{10}$ cm$^{-2}$V$^{-1}$, in reasonable agreement with that obtained from geometric considerations.

Interestingly, for large $B$>8T, we observe doubling of the oscillation frequency. This can be seen in the line traces in Fig. 3b-c, and in Fig. 3d that plots the high field portion of Fig. 3a, where the additional periods are indicated by arrows. Such doubling in frequency most likely arises from Zeeman splitting. At $B$=12T, its disappearance between 3K and 4.5K (see Fig. 4a) provides an upper bound for the Zeeman energy $g\mu_B B$, where $g$ is the g-factor and $\mu_B$ Bohr magneton. Using the simple estimate $g\mu_B B \sim k_B T$, we obtain $g$~1.8 to 2.7, which is reasonable.

Finally, we seek to measure the cyclotron mass of the charge carriers by investigating the temperature dependence of the oscillations. Fig. 4a presents $\Delta R_{xx}(V_g)$ at $B$=12T and different temperatures between 1.5K and 12K. The additional, Zeeman-induced oscillations disappear at $T$>~4K, and the main oscillations at $T$>15K. For a single period, the amplitude of the main oscillation is measured by taking the average of the height between the peak and the two adjacent troughs. To extract $m^*$, we fit the amplitude $A$ to the temperature-dependent terms of the Lifshitz-Kosevich formula at constant $E_F$

$$A(T) = \frac{CT}{\sinh(bT)} \qquad (2)$$

where $C$ and $b = \frac{2\pi^2 k_B m^*}{\hbar e}$ are fitting parameters (Fig. 4b). Reasonable agreement with data points are obtained, yielding $m^*$ measured at different $V_g$ values. As shown in Fig. 4c, $m^* \approx 0.31 m_e$ ($m_e$ is the rest mass of electrons) at $V_g$=-30V or $n \approx$-4.6x10$^{12}$ cm$^{-2}$. As $V_g$ decreases to -64V ($n \approx$-7.4x10$^{12}$ cm$^{-2}$), $m^*$ becomes lighter ~0.25$m_e$.

Theoretically, since the energy dispersion at the band minima in FLP is anisotropic, the effective masses along different principal axes are dramatically different. The cyclotron mass extracted from SdH oscillations is the geometric mean of those along different axes in the $x$-$y$ plane, $m^* = \sqrt{m_x^* m_y^*}$. Using density functional theory (DFT) implemented in the Vienna Ab-initio Simulation Package (VASP), we calculate the valence band effective masses along the $k_x$ and $k_y$ direction for the experimentally explored range of density for FLP with different thicknesses. Our results show that as $n$ decreases and $E_F$ moves towards the band edge, $m_x^*$ remains fairly constant, 0.11-0.12$m_e$ for all thickness; in contrast, $m_y^*$ exhibits strong dependence on the number of layers and on $E_F$. For instance, at $n$=-4.8x10$^{12}$ cm$^{-2}$, $m_y^*/m_e$=6.2 for monolayer phosphorene, and 2.0 for 25-layer phosphorene; at $n$=-8.3x10$^{12}$ cm$^{-2}$, these values decreases to 4.4 and 1.2, respectively. Fig. 4d shows the theoretically calculated cyclotron mass $\sqrt{m_x^* m_y^*}$ as a function of $V_g$. All FLP of different thicknesses exhibit the general trend of an increase in $m^*$ towards the band edge, as observed experimentally. The theoretically calculated values of $m^*$ agree with the experimentally measured values within 50%, which is reasonable.

In conclusion, we demonstrate that hBN/phosphorene/hBN heterostructures with 1D edge contacts enable exploration of air-stable single- and few-layer phosphorene devices, with mobility up to 4000 cm$^2$/Vs. As temperature decreases, we observe a gate-tunable metal-insulator transition. At low temperatures and moderate magnetic fields, prominent SdH oscillations establish the presence of Zeeman-split Landau levels within the sample, and enable experimental determination of the cyclotron mass of charge carriers as the Fermi level is tuned by a gate voltage. Our work opens the door to synthesis of stable ultra-high mobility devices based on phosphorene and other 2D semiconductors, thus providing exciting platforms for the investigation of fundamental 2D processes in reduced dimensions and for electronic and optoelectronic applications.

The work is supported by the FAME center, one of the six STARnet centers supported by DARPA and SRC. YS is supported in part by ONR. YB is partially supported by CONSEPT center at UCR. JH is supported by NSF/ LA-SiGMA program under award #EPS-1003897. ZQM acknowledges the support from NSF under Grant No. DMR-1205469. This work used the Extreme Science and Engineering Discovery Environment (XSEDE), which is supported by National Science Foundation grant number OCI-1053575.

**Methods**
**Synthesis of Bulk Black Phosphorus**
The BP single crystal was synthesized using a chemical vapor transport method modified from that of the earlier reports[23]. A mixture of red phosphorus, AuSn, and SnI$_4$ powder with mole ratio 1000:100:1 was sealed into an evacuated quartz tube. The tube is then placed into a double-zone tube furnace with temperature set at 600°C and 500 °C for the hot and cold end, respectively. Large single crystals can be obtained after a weeks of transport.

**Ab initio Calculations**
*Ab-initio* calculations were used to calculate the valence band effective masses of bulk and few-layer black phosphorus structures. Density functional theory (DFT) with a projector augmented wave method[35] and the Perdew-Burke-Ernzerhof (PBE)[36] type generalized gradient approximation as implemented in the Vienna Ab-initio Simulation Package (VASP) [37, 38]was used. The van-der-Waal interactions in black phosphorus were accounted for using a semi-empirical correction to the Kohn-Sham energies when optimizing the bulk structure[39]. The lattice parameters of the monolayer and the few-layer structures are *a*=4.592Å and *b*=3.329Å along the armchair and the zig-zag directions respectively. The energy cutoff of the plane wave basis is 500 eV. A Monkhorst-Pack scheme is used to integrate over the Brillouin zone with a k-mesh of (16x16x8) and (16x16x1) for the bulk and few-layer structures respectively. To verify the results of the PBE band structure calculations of bulk and one to four layers of black phosphorus were calculated using the Heyd-Scuseria-Ernzerhof (HSE) functional[40]. The HSE calculations incorporate 25% short-range Hartree-Fock exchange. The screening parameter μ is set to 0.2 Å-1. The effective masses along the arm-chair ($m_x$) and zig-zag ($m_y$) directions are obtained by fitting the energy dispersion to an even sixth order polynomial. For each structure, the valence band effective masses along $k_x$ are calculated from 0.04 (2π/*a*) to 0.06 (2π/*a*) and along $k_y$ from 0.08 (2π/*a*) to 0.09 (2π/*a*) where *a* is the lattice constant along the armchair direction. This corresponds to varying the hole density from 4.8x10$^{12}$ cm$^{-2}$ to 8.3x10$^{12}$ cm$^{-2}$.

Fig. 1. (a). Atomic configuration of monolayer phosphorene. (b). Schematic of fabrication process. hBN/FLP/hBN stacks are created via successive dry transfer techniques, etched to expose the edges of phosphorene, then coupled to Cr/Au electrodes via one-dimensional edge contacts. (c). False-color optical microscope image of a completed device. Inset: schematic of the device's side view. (d). Two-terminal conductance $G$ of a hBN/FLP/hBN heterostructure *vs.* gate voltage $V_g$. The different traces correspond to data taken successively after different hours of exposure to ambient conditions.

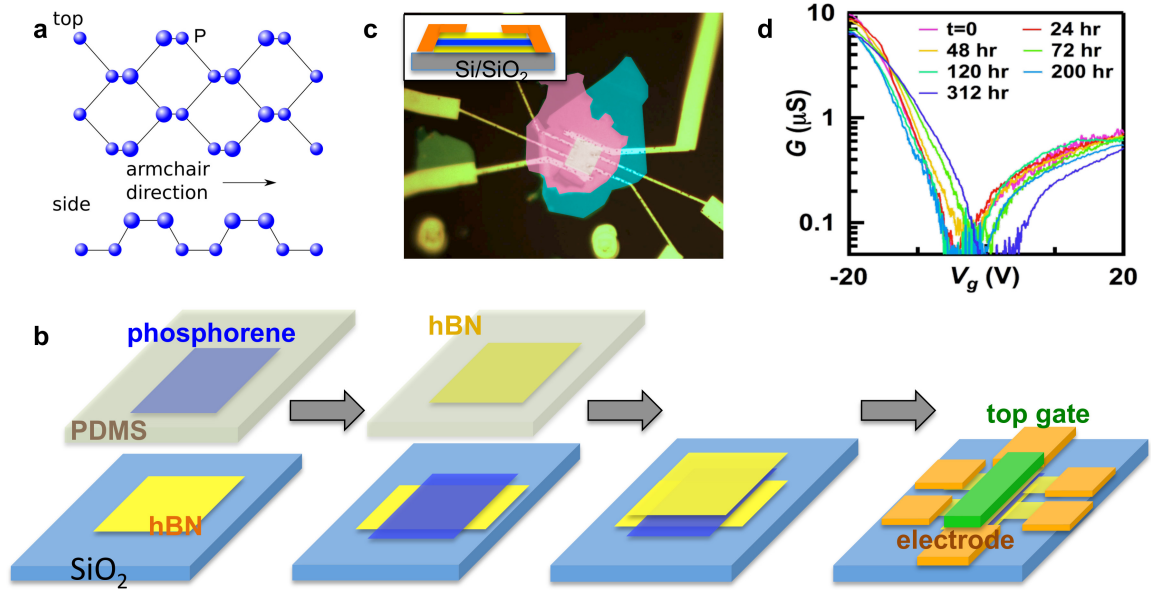

Fig. 2. Transport data at $B=0$. (a-b). $G(V_g)$ of a 10-nm-thick hBN/FLP/hBN heterostructure at $T=300$K and 1.5K, respectively. The two curves in (a) correspond to different sweeping direction. (c). Current-voltage characteristics at $T=1.6$K and different gate voltages. (d). Four-terminal resistance $R_{xx}$ vs. $V_g$ at different temperatures. (e). $R(T)$ at $V_g=$ 70, -50, -30, -25, -20, -17 and -15V, respectively (bottom to top). (f). Mobility $\mu(T)$ for $V_g=-70, -25$ and $-15$V.

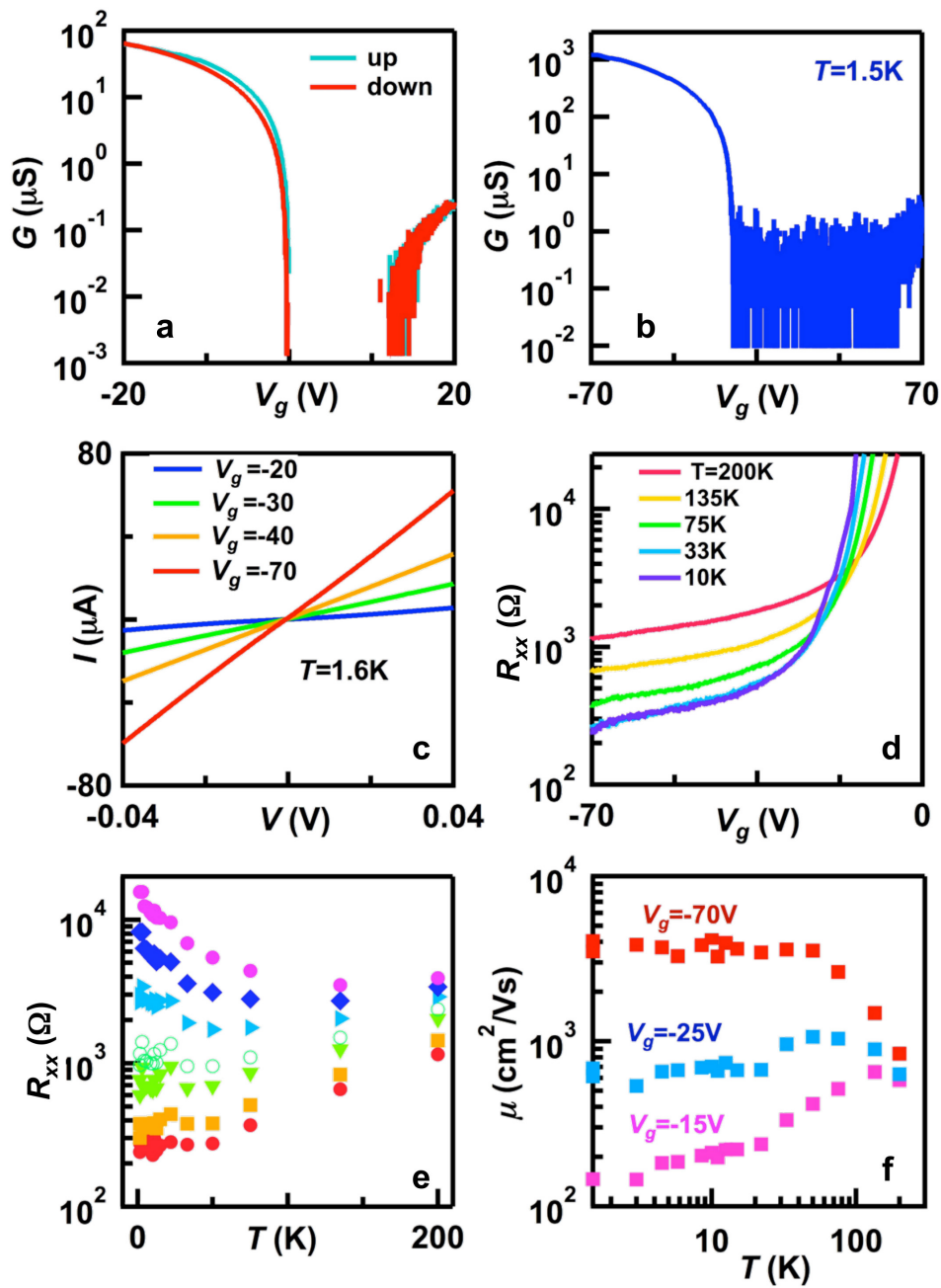

Fig. 3. (a). Oscillations $\Delta R_{xx}$ (color) vs. $V_g$ and $B$. A smooth background is subtracted from the resistance data. (b). $\Delta R_{xx}(V_g)$ at different magnetic fields. The traces are offset for clarity. (c). $\Delta R_{xx}$ vs. $B$ (left) and $1/B$ (right) at different $V_g$. The traces are offset for clarity. (d). A zoom-in plot of the oscillations $\Delta R_{xx}(V_g, B)$ in high fields. The arrows indicate the appearance of the second period induced by Zeeman splitting.

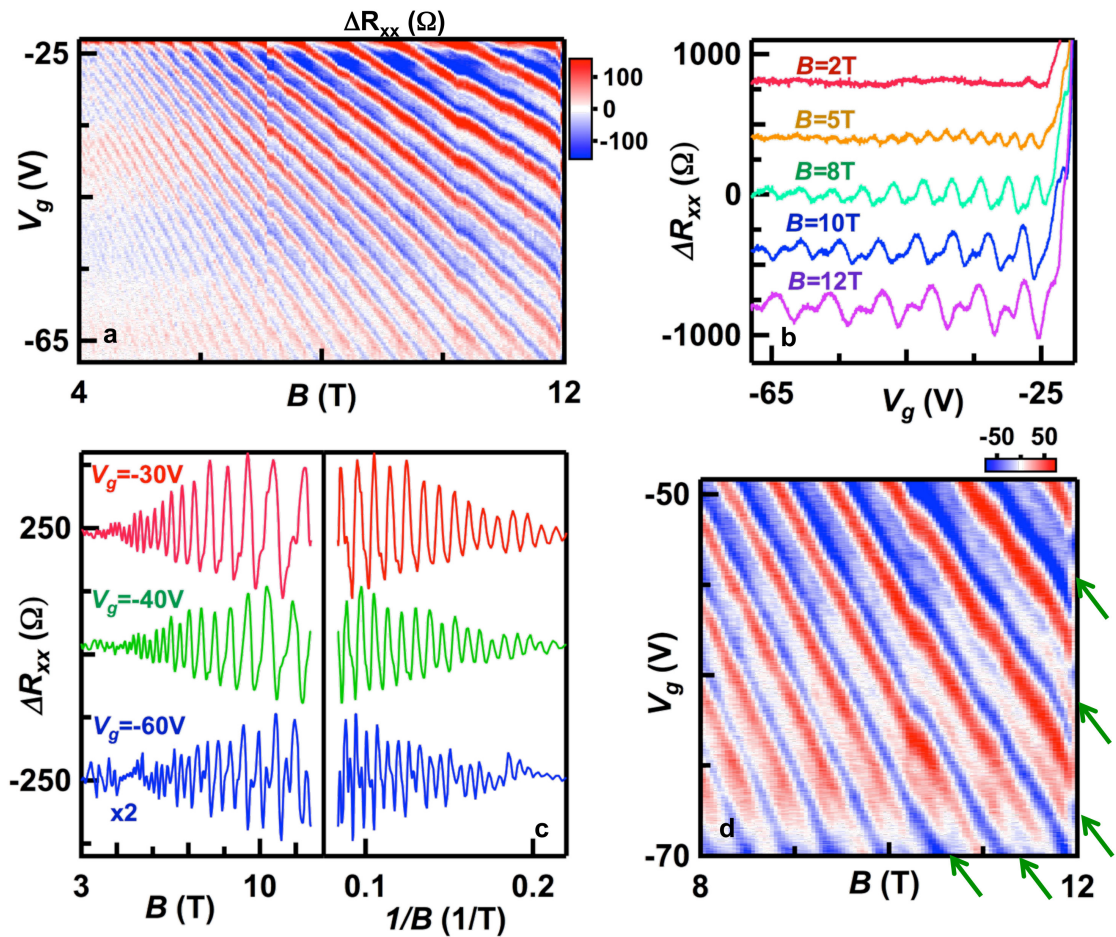

Fig. 4. (a). $\Delta R_{xx}(V_g)$ taken at $B$=12T and $T$=1.5, 3, 4.5, 5.8, 7, 8.5, 10, 11, 12.5 and 15K, respectively (bottom to top). (b). Data points are measured oscillation amplitude vs. $T$ for the peaks at $V_g$= -30, -36, -43, -50, -57, -64V, respectively (bottom to top). The lines are fits to Eq. (2). The traces in (a) and (b) are offset for clarity. (c). Extracted cyclotron mass from SdH oscillations as a function of $V_g$. (d). Cyclotron masse $\sqrt{m_x^* m_y^*}$ from DFT calculations for FLP with different number of layers.

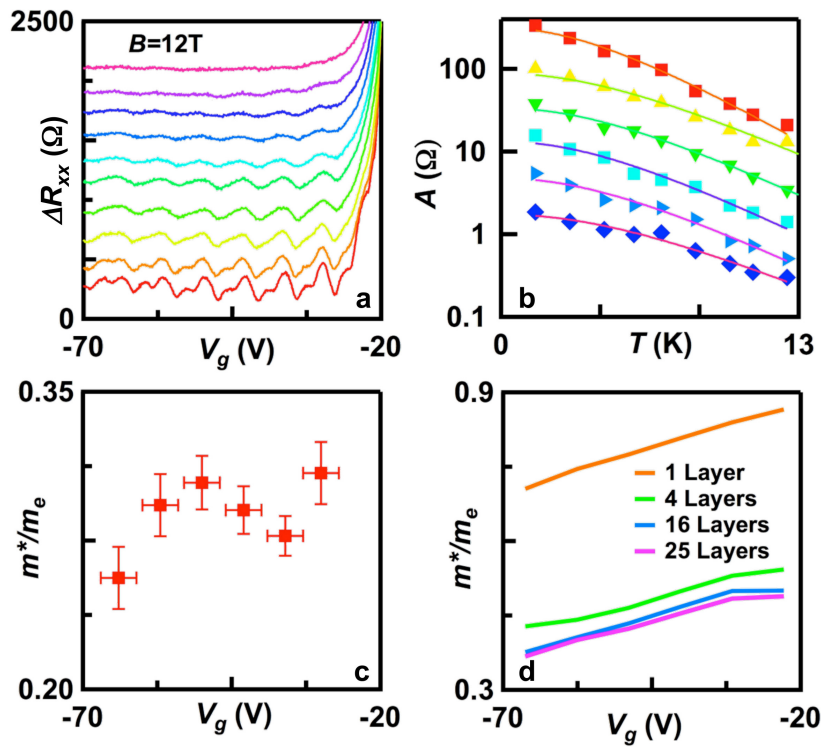